\documentclass[onecolumn]{emulateapj}
\usepackage{scalerel}
\usepackage{tikz}
\usetikzlibrary{svg.path}

\newcommand{\orcid}[1]{\href{https://orcid.org/#1}{\textcolor[HTML]{A6CE39}{\aiOrcid}}}

\definecolor{orcidlogocol}{HTML}{A6CE39}
\tikzset{
    orcidlogo/.pic={
        \fill[orcidlogocol] svg{M256,128c0,70.7-57.3,128-128,128C57.3,256,0,198.7,0,128C0,57.3,57.3,0,128,0C198.7,0,256,57.3,256,128z};
        \fill[white] svg{M86.3,186.2H70.9V79.1h15.4v48.4V186.2z}
        svg{M108.9,79.1h41.6c39.6,0,57,28.3,57,53.6c0,27.5-21.5,53.6-56.8,53.6h-41.8V79.1z M124.3,172.4h24.5c34.9,0,42.9-26.5,42.9-39.7c0-21.5-13.7-39.7-43.7-39.7h-23.7V172.4z}
        svg{M88.7,56.8c0,5.5-4.5,10.1-10.1,10.1c-5.6,0-10.1-4.6-10.1-10.1c0-5.6,4.5-10.1,10.1-10.1C84.2,46.7,88.7,51.3,88.7,56.8z};
    }
}

\newcommand\orcidicon[1]{\href{https://orcid.org/#1}{\mbox{\scalerel*{
                \begin{tikzpicture}[yscale=-1,transform shape]
                \pic{orcidlogo};
                \end{tikzpicture}
            }{|}}}}

\usepackage{amssymb}
\usepackage{amsmath}
\usepackage{graphicx}
\usepackage{dcolumn}
\usepackage{hyperref}
\usepackage{ulem}
\hypersetup{
    colorlinks=true,
    citecolor=blue,
    linkcolor=blue,
    urlcolor=blue,
}

\shorttitle{Removing ZTF Atmospheric Fringes} 
\shortauthors{Medford et al.}

\begin{document}

\title{Removing Atmospheric Fringes from Zwicky Transient Facility i-Band Images using Principal Component Analysis}

\author{\orcidicon{0000-0002-7226-0659} Michael S. Medford}
\affiliation{Department of Astronomy, University of California, Berkeley, Berkeley, CA 94720}
\affiliation{Lawrence Berkeley National Laboratory, 1 Cyclotron Rd., Berkeley, CA 94720}

\author{\orcidicon{0000-0002-3389-0586} Peter Nugent}
\affiliation{Lawrence Berkeley National Laboratory, 1 Cyclotron Rd., Berkeley, CA 94720}

\author{\orcidicon{0000-0003-3461-8661} Danny Goldstein}
\affiliation{Lawrence Berkeley National Laboratory, 1 Cyclotron Rd., Berkeley, CA 94720}

\author{\orcidicon{0000-0002-8532-9395} Frank J. Masci}
\affiliation{IPAC, California Institute of Technology, 1200 E. California
             Blvd, Pasadena, CA 91125, USA}
             
\author{\orcidicon{0000-0003-3768-7515} Igor Andreoni}
\affiliation{Division of Physics, Mathematics, and Astronomy, California Institute of Technology, Pasadena, CA 91125, USA}
             
\author{Ron Beck}
\affiliation{IPAC, California Institute of Technology, 1200 E. California
             Blvd, Pasadena, CA 91125, USA}
             
\author{\orcidicon{0000-0002-8262-2924} Michael W. Coughlin}
\affil{School of Physics and Astronomy, University of Minnesota,
Minneapolis, Minnesota 55455, USA}

\author{\orcidicon{0000-0001-5060-8733} Dmitry A. Duev} 
\affiliation{Division of Physics, Mathematics, and Astronomy, California Institute of Technology, Pasadena, CA 91125, USA}

\author{\orcidicon{0000-0003-2242-0244} Ashish~A.~Mahabal}
\affiliation{Division of Physics, Mathematics and Astronomy, California Institute of Technology, Pasadena, CA 91125, USA}
\affiliation{Center for Data Driven Discovery, California Institute of Technology, Pasadena, CA 91125, USA}

\author{\orcidicon{0000-0002-0387-370X} Reed L . Riddle} 
\affil{Caltech Optical Observatories, California Institute of Technology, Pasadena, CA 91125, USA}

\begin{abstract}
The Zwicky Transient Facility is a time-domain optical survey that has substantially increased our ability to observe and construct massive catalogs of astronomical objects by use of its 47 square degree camera that can observe in multiple filters.
However the telescope's i-band filter suffers from significant atmospheric fringes that reduce photometric precision, especially for faint sources and in multi-epoch co-additions.
Here we present a method for constructing models of these atmospheric fringes using Principal Component Analysis that can be used to identify and remove these artifacts from contaminated images.
In addition, we present the Uniform Background Indicator as a quantitative measurement of the reduced correlated background noise and photometric error present after removing fringes.
We conclude by evaluating the effect of our method on measuring faint sources through the injection and recovery of artificial stars in both single-image epochs and co-additions.
Our method for constructing atmospheric fringe models and applying those models to produce cleaned images is available for public download in the open source python package \href{https://github.com/MichaelMedford/fringez}{fringez}.
\end{abstract}

% \maketitle

\section{Introduction}

Large scale synoptic surveys have produced an abundance of optical images at scales previously unseen.
Surveys such as the Palomar Transient Factory \citep{Law2009}, Catalina Real-Time Transient Survey \citep{Drake2009}, Zwicky Transient Facility \citep{Bellm2019a, Graham2019, Masci2018} and others have revolutionized our understanding of the universe by generating massive transient datasets.
These datasets require advances in computational processing techniques.
The Vera C. Rubin Observatory will generate 20 terabytes of data per night, totalling over 500 petabytes of imaging data over the 10 years of the survey \citep{rubin}.
Such data flows require that the reduction of raw images into calibrated science images must be systematic and require minimal human intervention.

One significant source of noise in long wavelength optical imaging is atmospheric emission lines.
These emission lines are produced by highly non-thermal atomic and molecular transitions (primarily \ion{O}{2} and OH) and are influenced by the temperature and density of the upper and lower atmosphere as well as the current solar activity.
Thus the strength of these lines can vary throughout the night and is proportional to the airmass through which the telescope is pointed.
Fringe patterns appear when photons at these wavelengths fall onto thin charge-coupled devices (CCDs) due to the self-interference caused from light reflecting off of the back of the imaging instrument before it is absorbed by the CCD itself \citep{Bernstein_2017}.
Thick CCDs rarely see this effect save at the longest wavelengths.
For thin CCDs, these fringe patterns can introduce significant noise into i-band and z-band images, rendering photometry and image subtraction ineffective at faint magnitudes without a calibration correction that successfully removes them.
These fringes also appear in interference image spectrometers where attempts have been made to remove them using wavelet transformations \citep{Ren:17}.

Principal component analysis (PCA) is a statistical method for reducing data to a set of orthogonal components by finding vectors of minimal variance within the data \citep{Jolliffe2016}.
PCA has been shown to be an effective method for modeling and removing atmospheric fringes from red optical and near-IR images.
These methods work by building a set of orthogonal component images, constructed from a large representative sample, which can approximate the fringes of a single image through linear combination.
The dot product of individual images against these orthogonal components results in eigenvalues that, when used as weights to these orthogonal images, generate bias images of the atmospheric fringes.
Subtracting these bias images removes the fringe noise and improves the photometric precision of astrophysical source measurements.

Previous PCA methods for atmospheric fringe subtraction constructed orthogonal images by down-sampling fringed images into a lower resolution before re-parameterization of the data.
\cite{Bernstein_2017} compress an image into a sparse set of features before attempting reduction into orthogonal components.
This method has strong results but lacks the ability to resolve contributions from individual atmospheric lines due to its compression.
PCA performed on a per-pixel basis is more computationally expensive but has the power to capture these individual fringe effects, resulting in more accurate photometry.

The computational resources dedicated to the reduction and calibration of astronomical images has grown over the decades to keep up with increasing data flows.
The development of highly optimized PCA algorithms, along with additional computational resources, have now made it possible to model fringes via per-pixel PCA analysis.
Here we present the implementation of such a method on full resolution Zwicky Transient Facility i-band data.
In section \ref{sec:telescope}, we outline the Zwicky Transient Facility instrument and dataset.
In Section \ref{sec:method}, we present our method for implementing per-pixel PCA atmospheric fringe modeling and removal, as well as the Uniform Background Indicator as a quantitative measurement of correlated background noise.
In Section \ref{sec:results}, we analyze the results of applying our method, including increased photometric precision on faint sources and the ability to detect otherwise undetectable sources in multi-epoch co-additions.
We discuss and conclude in Section \ref{sec:conclusion}.

\section{The Zwicky Transient Facility Instrument} \label{sec:telescope}

\nocite{Medford2020}

The Zwicky Transient Facility (ZTF) is an optical time-domain survey that has been operating on the 48-inch Samuel Oschin Telescope at Palomar Observatory since March 2018 \citep{Bellm2019a, Graham2019}.
ZTF's camera covers 47 square degrees in a single exposure, enabling coverage of the entire visible Northern sky every few nights in ZTF g-band, r-band and i-band filters with an average point spread function (PSF) full width at half maximum of $2.0''$ on a plate scale of $1.01''\ \text{pixel}^{-1}$.
The ZTF camera is divided into 16 CCDs, each covered with an anti-reflective (AR) coating, with each CCD split into four separate readout channels for a total of 64 readout channels \citep{Dekany2020}.
Surveys with the telescope over its first several years of operations \citep{Bellm2019b} take standard 30 second exposures that achieve median five-sigma limiting magnitudes of $r\approx21.0$, $g\approx21.25$, and $i\approx20.0$ \citep{Bellm2019a}.
The Infrared Processing and Analysis Center (IPAC) reduces the data for the ZTF survey, including producing a real-time alert stream triggered by transient event detections on difference images \citep{Patterson2018}.
In addition to these alerts, the ZTF collaboration routinely produces public data releases that contain, among other data products, lightcurves assembled from single image PSF photometry for every star in the northern sky that appears in a deep co-added reference image \citep{Masci2018}.
Reference images are ideally constructed from 40 individual exposures, although weather and visibility produces variable results for different areas of the sky.
ZTF's observing time is split between public observations (40\%) (funded by the National Science Foundation's Mid-Scale Innovations Program or MSIP), partnership observations (40\%) (which are held in a proprietary period for collaboration members of the survey), and Caltech observations (20\%).
Starting in 2020, ZTF began operating its second phase as ZTF-II with 50\% of observation time funded by a MSIP survey that images the entire visible sky every other night.
The ZTF i-band filter is used exclusively for partnership observations, such as the ZTF Ultra Deep Survey (ZUDS), {\bf before being released to the public after the partnership's proprietary period}.
Due to the decreased limiting magnitude of the i-band relative to the g- and r-band, standard exposures in the i-band are set to 90 seconds and increase the filter's median limiting magnitude to a comparable value.
Atmospheric fringes appear exclusively in the i-band and therefore this work uses partnership data not yet released to the public.

\section{Method for Removing Atmospheric Fringes} \label{sec:method}

\begin{figure}[!htb]
    \centering
    \includegraphics[width=0.485\textwidth, angle=0]{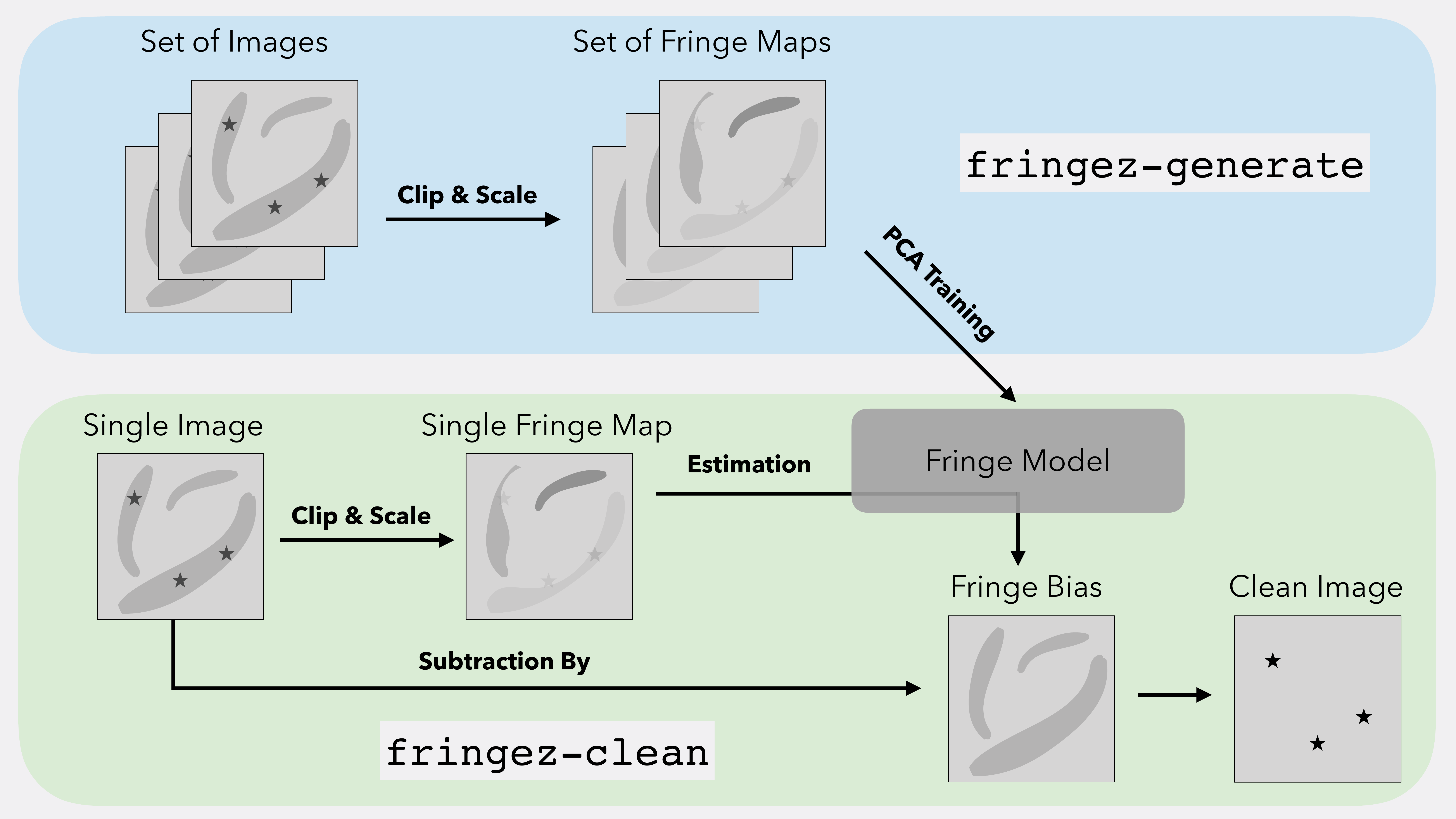}
    \caption{The process for removing atmospheric fringes was done in two steps. A fringe model for each readout channel was constructed by running fringe maps through PCA training, where fringe maps are a clipped and scaled version of single-epoch images. This step was only performed once per readout channel. Cleaning every single-epoch image thereafter was performed by using the eigen-vectors in the fringe model to generate a fringe bias image for each fringed image. This fringe bias was subtracted from the fringed image to create a clean image. Functions for generating fringe models and cleaning fringed images can be found in the open source  \href{https://github.com/MichaelMedford/fringez}{fringez} package under the executables \texttt{fringez-generate} and \texttt{fringez-clean} respectively. \label{fig:fringez_diagram}}
\end{figure}

Our method for removing atmospheric fringes was a two step process.
First the PCA eigen-vectors for a single readout-channel were extracted from a large set of images with significant fringing.
Second each image was processed through this model to generate a unique bias image that, when subtracted from the original image, removed the atmospheric fringes.
Figure \ref{fig:fringez_diagram} outlines a visual representation of the steps in this method.
{\bf An explanation of the data products and corresponding symbols used throughout this section can be found in Table \ref{tab:fringez_symbols}.}

\begin{table}[t]
  \centering
  \caption{\texttt{fringez} Math Symbols}
    \begin{tabular}{ccc}
      Description & Symbol & Dimensions \\
      \hline
      Fringed Image & $\mathbf{I}_\text{fringe}$ & $[N_\text{rows} \times N_\text{cols}]$ \\ 
      Flattened Fringed Image & $\overrightarrow{I}_\text{fringe}$ & $[1 \times N_\text{pixels}]$ \\ 
      Fringe Map & $\overrightarrow{F}_\text{map}$ & $[1 \times N_\text{pixels}]$ \\ 
      Fringe Map Eigen-values & $\overrightarrow{\lambda}_\text{map}$ & $[1 \times N_\text{comp}]$ \\
      Fringe Bias & $\overrightarrow{F}_\text{bias}$ & $[1 \times N_\text{pixels}]$ \\ 
      Average of Training Set & $\overrightarrow{\mu}_\text{model}$ & $[1 \times N_\text{pixels}]$ \\
      Model Eigen-images & $\mathbf{v}_\text{model}$ & $[N_\text{comp} \times N_\text{pixels}]$ \\
      Model Explained Variance & $\overrightarrow{R}_\text{model}$ & $[1 \times N_\text{pixels}]$ \\
    \end{tabular}
    \label{tab:fringez_symbols}
\end{table}

\subsection{Generating PCA Fringe Models} \label{sec:method_model}

ZTF's reduction pipeline was designed to process each of the camera's 64 readout-channels separately.
We therefore began by gathering images of a common readout-channel together for model generation, with the goal of generating 64 fringe models.
Training images were selected by gathering all i-band images between 2019-04-01 and 2020-04-01 and removing images with a limiting magnitude less than 19.
This cut on limiting magnitude removed cloudy images from our sample that would fail to have a photometrically accurate measurement of the night sky and thus its atmospheric lines.
By including all images within these dates we ensured a representative sample of atmospheric conditions and airmasses which are correlated with the strength of atmospheric fringes.
There were 550,365 i-band images included in total, with each fringe model trained on between 8,062 and 8,898 images.

We began by selecting all fringed images ($\mathbf{I}_\text{fringe}$) for a single readout channel.
The location and strength of atmospheric fringes are independent of the astrophysical sources within an image.
Therefore we needed to remove sources from each image prior to training in order to reduce the confusion in the PCA variance reduction process.
Our preferred method for identifying pixels containing astrophysical sources was to use the quality masks produced by IPAC that labels pixels containing a source in each image's source catalog.
In cases where this pixel mask could not be obtained, pixels containing sources were identified by calculating the median absolute deviation for the image and flagging pixels that were 5 standard deviations above or below the image's median value.
{\bf This process did not clip the atmospheric fringes themselves, which were at the level of the background noise and therefore significantly below the level of these thresholds.}
All pixels flagged as containing sources were replaced with the value of the image's global median, removing stars from our images without changing the value of the vast majority of pixels.
{\bf A possible improvement to our method could be to replace the pixel instead with an estimate of the local background.
We choose the global median because the local background can be difficult to estimate due to the fringes themselves and in particular in regions where the transverse length scale of the fringes approaches the pixel scale of the image.}
Each image was then scaled so that they could be used for training together with other images taken at different airmasses and limiting magnitudes.
We named each training image a fringe map ($\overrightarrow{F}_\text{map}$), as it traces the relative location and strength of only the atmospheric fringes.
Fringe maps were created by subtracting the median from each image, dividing the result by the image's median absolute deviation, and flattening the image into a $[1 \times N_\text{pixels}]$ array:
\begin{align}
    \overrightarrow{F}_\text{map} &= \left.\frac{\mathbf{I}_\text{fringe} - \widetilde{\mathbf{I}}_\text{fringe}}{\text{MAD}(\mathbf{I}_\text{fringe})} \right\vert_\text{flattened}. \label{eq:fringe_map}
\end{align}
\sout{A reasonable assumption would be that collapsing the spatial correlation present in the 2-dimensional image would make our method less effective.
However this concern proved unfounded, as experimentation with fitting sub-images of different sizes resulted in no improvement on our method.}
{\bf (Because PCA is a linear method that calculates coefficients irrespective of the basis in which the data is represented, performing our analysis in one dimension has no effect on the results.)}
Each pixel in the fringe map was treated as a separate \sout{feature} {\bf variable} in our PCA analysis, resulting in 9,461,760 \sout{features} {\bf variables} (3080 rows by 3072 columns) for each model.
The eigenvalues \sout{for each feature} {\bf of the feature vector} was determined using the randomized Singular Value Decomposition method \citep{Martinsson_2011, Halko_2011} in the \texttt{scikit-learn} python package \citep{scikit-learn}.
Each 1-dimensional eigen-vector was reconstructed into the original image shape to create an eigen-image.
The final fringe model was a set of eigen-images that captured orthogonal contributions to the training set variance in each pixel.
{
\bf
We also recorded each component's explained variance ($\overrightarrow{R}_\text{model}$), or the amount of training set variance pointed in each eigen-vector's direction, for later use in the processing pipeline.
We then repeated this process to construct a unique fringe model for each of the camera's 64 readout channels.
}

\subsection{Removing Fringes from Single-Epoch Images} \label{sec:method_clean}

{
\bf
Fringes were removed from images by applying the fringe model in these steps:
}
\begin{align}
    \overrightarrow{\lambda}_\text{map} &= \frac{(\overrightarrow{F}_\text{map} - \overrightarrow{\mu}_\text{model}) \boldsymbol{\cdot} \mathbf{v}_\text{model}^T}{\sqrt{\overrightarrow{R}_\text{model}}}\ , \label{eq:coeffs} \\
    \begin{split}
    \overrightarrow{F}_\text{bias} &= \left[\left[\overrightarrow{\lambda}_\text{map}\boldsymbol{\cdot} \left(\sqrt{\overrightarrow{R}_\text{model}} \mathbf{v}_\text{model}\right)\right] + \right.\\ &\left.\overrightarrow{\mu}_\text{model}\right] \cdot \text{MAD}(\overrightarrow{I}_\text{fringe}) \label{eq:fringe_bias}\ , \end{split}\\
    \overrightarrow{I}_\text{clean} &= \overrightarrow{I}_\text{fringe} - \overrightarrow{F}_\text{bias}. \label{eq:image_clean}
\end{align}

{
\bf
First, a fringe map was regenerated for each fringed image in our sample.
The dot product of this fringe map (less the average value of each pixel in the training set $\overrightarrow{\mu}_\text{model}$) and the model's eigen-images was calculated, and then divided by the square root of that component's explained variance (Equation \ref{eq:coeffs}).
This produced the eigen-values ($\overrightarrow{\lambda}_\text{map}$) for each fringe map. 
These eigen-values were used as coefficients to linearly combine re-scaled eigen-images and the average values of the training sets were added back in (Equation \ref{eq:fringe_bias}).
This image, after being re-scaled by the median absolute deviation of the original fringed image, was named a fringe bias ($\overrightarrow{F}_\text{bias}$).
When this fringe bias was subtracted from the original fringed image it produced a clean image ($\overrightarrow{I}_\text{clean}$) with significantly less fringing (Equation \ref{eq:image_clean}).
}

\begin{figure}[!htb]
    \centering
    \includegraphics[width=0.48\textwidth, angle=0]{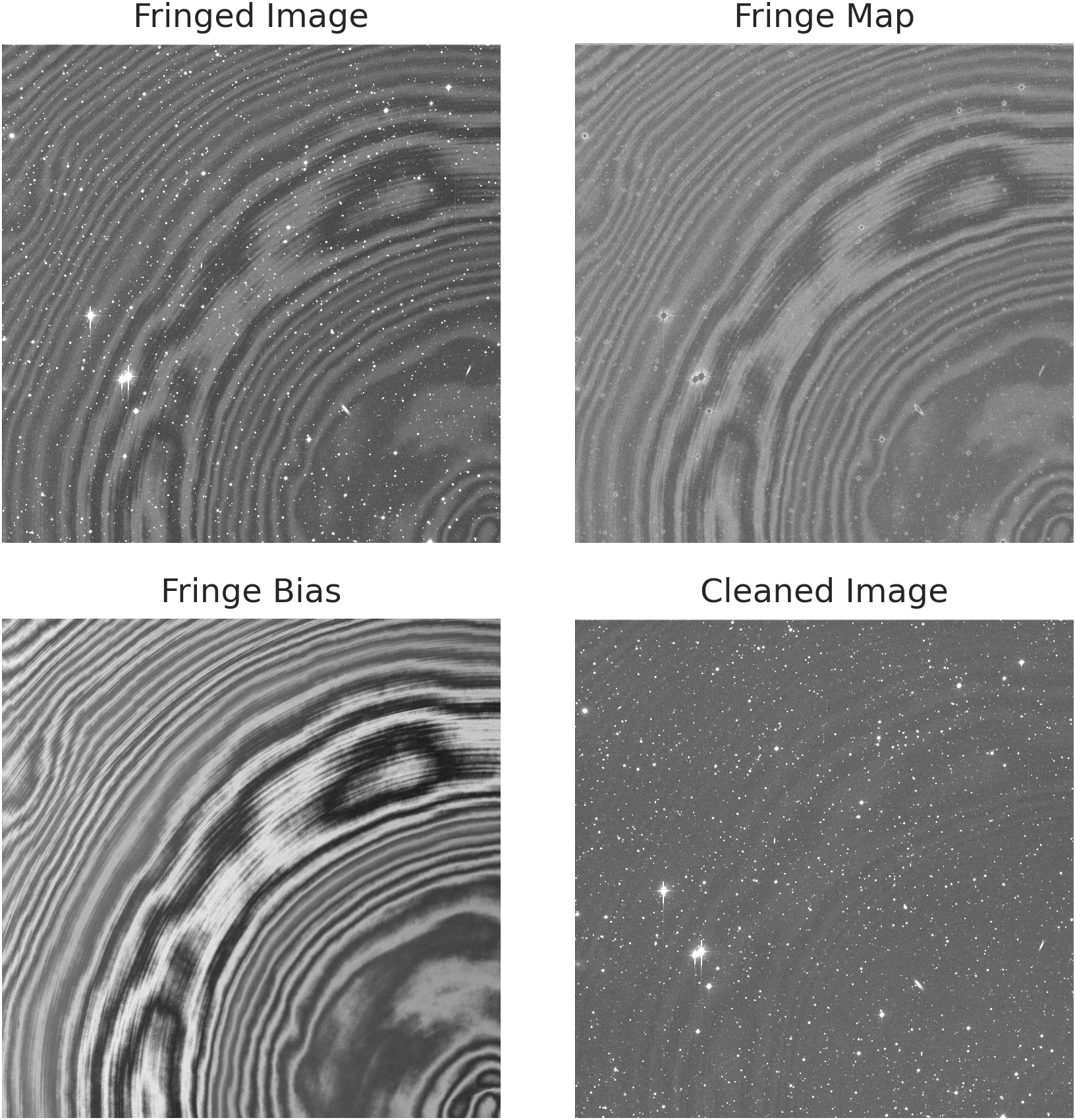}
    \caption{An example of an i-band image in the various steps of our fringe removal method. Fringed images were clipped and scaled to produce a fringe map. The fringe map was processed through a fringe model to generate a fringe bias. This bias image was subtracted from the original fringed image to create a cleaned image. While a fringe model only needs to be constructed once, each single-epoch image has a unique fringe bias determined by the linear combination of eigen-images in the fringe model that best reconstructs the fringe map. This ensures that the fringe model can successfully remove fringes arising from a variety of airmasses and seeing conditions. \label{fig:images}}
\end{figure}

An example of each of the images in this process are shown in Figure \ref{fig:images}.
This figure visually demonstrates the results of our method.
The original image is a 90 second i-band exposure that contains a representative amount of atmospheric fringes.
The fringe map is almost entirely devoid of individual sources, although the source pixel identification method does struggle around particularly bright stars.
The fringe bias generated from processing the fringe map through the fringe model successfully identifies the location and strength of each atmospheric fringe.
The clean image has nearly all of the atmospheric fringing pattern removed while retaining nearly all of the astrophysical sources.

\begin{figure}[!htb]
    \centering
    \includegraphics[width=0.48\textwidth, angle=0]{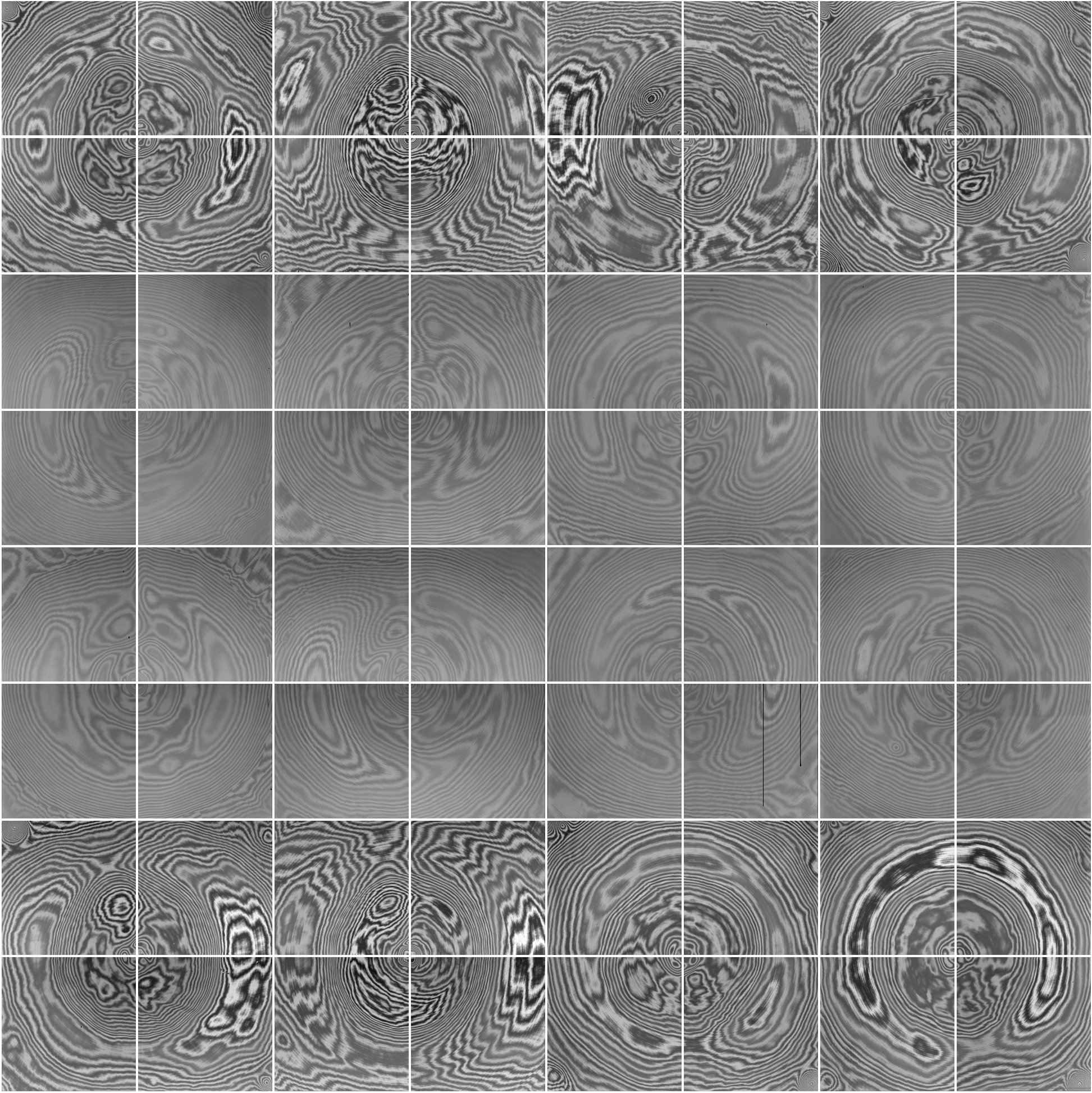}
    \caption{The average of the training fringe maps for each readout channel's fringe model on the ZTF camera, all placed onto a common gray-scale. These circular patterns roughly trace the thickness variations in the CCD and show the circular pattern resulting from flattening the device. The inner 32 readout channels show significantly less atmospheric fringing due to an additional layer of anti-reflective coating, as compared to the outer 32 readout channels. \label{fig:mosaic}}
\end{figure}

The arithmetic average of the fringe maps used to train the fringe model for each readout-channel is shown in Figure \ref{fig:mosaic}, with each readout-channel placed on a common gray-scale.
The pattern that clearly emerges is coherent on the level not just of the readout channel, but also over the full CCD.
The etching process for creating a thinned CCD uses a circular buffer that removes layers of the chip to create a uniformly thick device.
The average of the training fringe maps identifies the residual thickness variations for each CCD.
Work is ongoing to use these psuedo-measurements of the thickness to improve the ZTF data quality pipeline (Richard Dekany, private communication). 
The inner 32 readout channels ($16 \leq \text{rcid} < 48$) and the outer 32 readout channels ($0 \leq \text{rcid} < 16$, $48 \leq \text{rcid} < 64$) have distinctly different amounts of atmospheric fringing present in their images.
The inner readout channels have two layers of AR coating while the top and bottom rows of CCDs have a single layer coating.
This causes the outer readout channels to have a higher reflectivity at the longer wavelengths where fringing occurs and thus a larger contrast of the fringing pattern.

\begin{figure}[!htb]
    \centering
    \includegraphics[width=0.48\textwidth, angle=0]{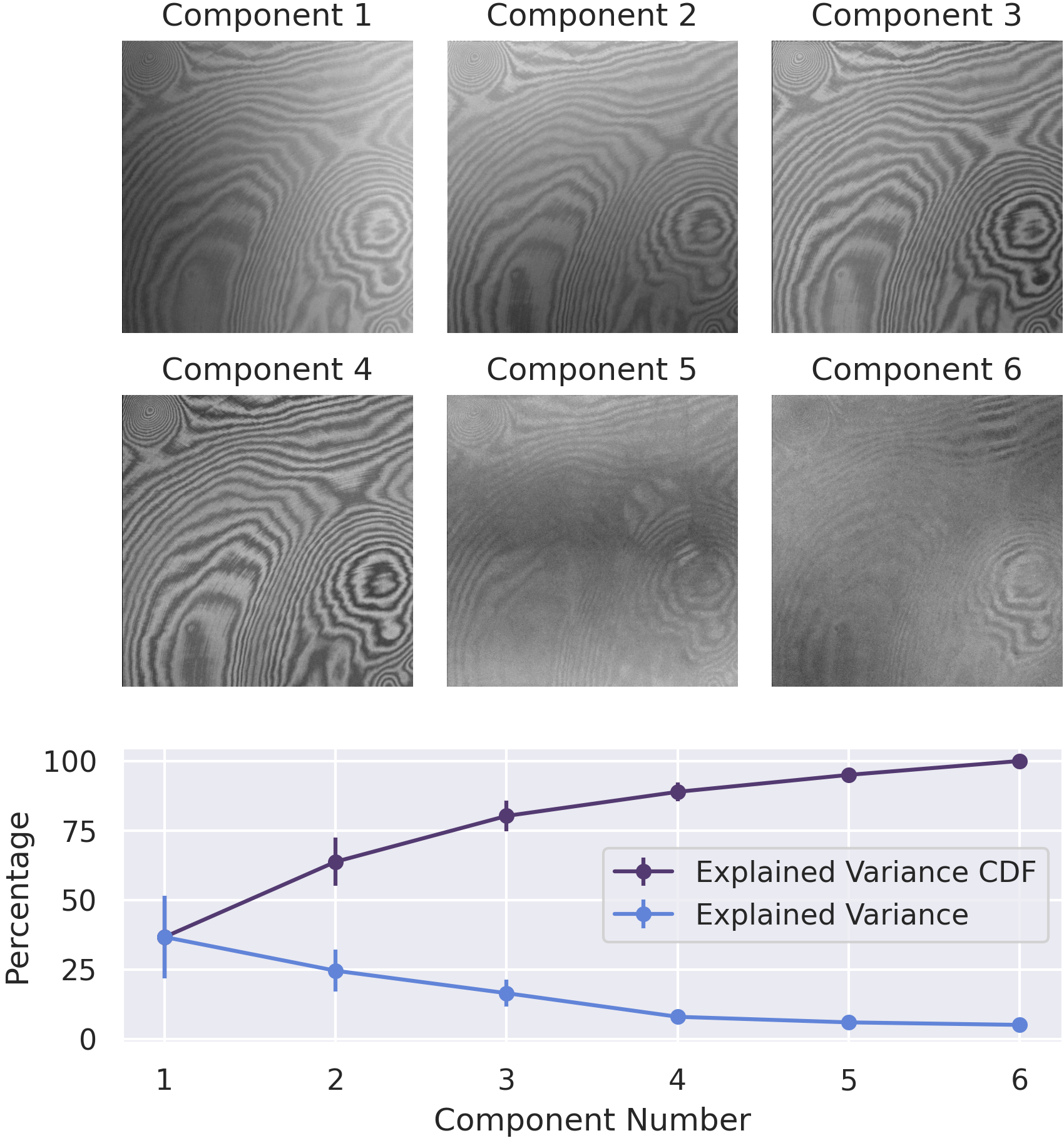}
    \caption{Each readout channel has a distinct fringe model with six eigen-images constructed from the reduction of thousands of training fringe maps. Here is an example of the eigen-images from readout channel 13 (top). While the fringe models were trained on a 1-dimensional array of pixels, the 2-dimensional fringes remain intact. The fringe pattern of the readout channel is clearly evident with slight variations in position and strength amongst the first four eigen-images. However the last two components (and to some extent the first four) contain smooth global gradients that would ideally be removed by flat fielding. The fractional explained variance for each component across all 64 fringe models (bottom) show that the fifth and sixth components captured far less variance in the training sample. The first component captured 64.6\% of the pixel variance while the sixth component captured only 5.0\% of the pixel variance. In total six PCA components reconstructs 95.0\% of the variance seen in our training sample. We therefore chose to have six components in our models. \label{fig:eigenimages}}
\end{figure}

The eigen-images for readout channel 13 are shown in Figure \ref{fig:eigenimages} as a representative example of a fringe model.
Each pixel in the image was trained as a separate \sout{feature} {\bf variable} and yet the fringe patterns across pixels remains coherent after reconstruction into the original image dimensions.
This confirms that the components contain correlated eigen-vectors for the different features.
In addition to the atmospheric fringes, the PCA training captured large scale variations in the background that remain after flat fielding.
These backgrounds dominate the higher order components.
This indicates that some global scale variations in flux remains after the execution of the ZTF flat-fielding pipeline that could possibly be improved by implementing a PCA method \citep{Bernstein_2017}.
We note that PCA requires setting the number of components, or number of eigen-vectors, for the reduction algorithm.
The fractional explained variance for all of the components are also shown in Figure \ref{fig:eigenimages}.
On average across the 64 readout channels, the first component captured 64.6\% of the pixel variance while the sixth component captured only 5.0\% of the pixel variance.
In total six PCA components reconstructed 95.0\% of the variance seen in our training sample.
Adding additional components failed to significantly increase the amount of fractional explained variance as the previous component.
We therefore determined that six components was sufficient to capture the variation due to atmospheric fringes.

Successfully training the fringe models required simultaneously holding large numbers of training images in computer memory in order to execute singular value decomposition.
The large number of images that trained each of our 64 fringe models ranged in memory from 285 gigabytes to 314 gigabytes. 
We trained our models on the Cori Haswell nodes at the NERSC Supercomputer located at the Lawrence Berkeley National Laboratory.
The Haswell nodes each have 128 gigabytes of RAM, making it impossible to include such a large number of images in our training sets without modifying our method.
To adapt to cases where access to computational resources are limited such as this one, our training method was modified by down-sampling the training set.
Training images were sorted by limiting magnitude and split into 300 sub-stacks of approximately equal depth.
{\bf
In order to combine each sub-stack into a single training image we calculated the sub-stack's median.
While taking the median can distort the correlations between pixels that PCA is attempting to solve for, the median is also more robust than the mean to removing outliers that have contaminated the signal from a different distribution than the one attempting to be measured.
Smooth background gradients remain on some ZTF images that can confuse the PCA reduction algorithm if not removed as outliers.
We therefore settled on taking the median instead of the mean to combine the sub-stack of fringe maps into a single training image.
An alternative approach could be to entirely remove the images with smooth gradients through a different method and then combine the remaining images in the sub-stacks using the mean.
}
\sout{A single training image was created from each sub-stack by taking the median of the fringe maps in the sub-stack.}
These 300 training images totalled approximately 11 GB and were therefore able to be fed simultaneously into our training process.
Comparing the fringe models that result from training on down-sampled fringe maps produced comparable results to training on individual fringe maps for smaller sample sizes.

The code used for generating fringe models from ZTF i-band images and cleaning fringed images has been released in the open-source package \texttt{fringez} \citep{fringez}.
The \texttt{fringez} package includes command line executables and python functions for downloading the fringe models computed as a result of this work, producing fringe bias images, and cleaning fringed images.
This package is available for download and installation at \href{https://github.com/MichaelMedford/fringez}{https://github.com/MichaelMedford/fringez}, and all fringe models are available for download at \href{https://portal.nersc.gov/project/ptf/iband}{https://portal.nersc.gov/project/ptf/iband}.
In October 2019, \texttt{fringez} was implemented into the ZTF IPAC data reduction pipeline with fringe models built on 500 training images per readout channel.
All i-band images taken previous to this date were also reprocessed by IPAC to remove atmospheric fringes.
In November 2020, fringe models were updated to include the 550,365 i-band images described and investigated in this paper.
This implementation of \texttt{fringez} and the current version of fringe models will continue to be a part of the IPAC data reduction pipeline in ZTF-II.
{\bf
The current implementation of \texttt{fringez} can also be generalized to other instruments by simply altering the FITS extensions and header keywords that are currently chosen for ZTF images.
}

\section{Measuring Improved Photometric Precision} \label{sec:results}

\subsection{The Uniform Background Indicator}

{
\bf
In order to assess the effectiveness of our de-fringing method, we created a procedure for measuring correlated background noise.
In this method, aperture photometry is taken at random locations on the background of an image.
At the location of each aperture the flux ($F_{\rm bkg}$) and its associated measurement error ($\sigma_{\rm bkg}$) are measured, resulting in a set of aperture fluxes ($\{F_{\rm bkg}\}$) and a set of errors ($\{\sigma_{\rm bkg}\}$).
We then compare the standard deviation of background fluxes ($\text{std}(\{F_{\rm bkg}\})$) to the average error on the background flux measurements ($\text{median}(\{\sigma_{\rm bkg}\})$).
}
We call the ratio of these two values the Uniform Background Indicator:

\begin{align} 
	\Psi &= \frac{\text{std}(\{F_{\rm bkg}\})}{\text{median}(\{\sigma_{\rm bkg}\})}. \label{eq:UBI}
\end{align}

For an image of uniform Gaussian noise, $\Psi\approx1$ as the distribution in background fluxes will be equal to the average flux error across all measurements.
Here there exists no global variance which is not captured by the local error term.
For an image with correlated background noise, $\Psi > 1$ because the different background values sampled by the apertures will introduce additional variance into the numerator of Equation~\ref{eq:UBI} not captured by the local error terms in the denominator.
{\bf The flux errors in Equation \ref{eq:UBI} must be calculated locally and be in the same units as the flux measurement with respect to aperture area.}
\sout{
The aperture flux errors must be calculated locally, for example by taking the root-mean-square of the median-subtracted pixel values in a small square around each pixel.
When calculating the flux error, the additional aperture area term that arises from combining pixel variances must be divided out so that both the flux and flux error have equal units.
The size of this square must be smaller than the spatial scale of the correlated background noise so that it contains minimal variation in the background value.
We found that setting the flux error equal to the root-mean-square of the median-subtracted pixel values of a 10 pixel by 10 pixel square around each pixel to be the right size for characterizing ZTF i-band atmospheric fringe patterns.}
The aperture size should be chosen to be approximately the PSF scale of the instrument \textbf{so that UBI measurements will indicate what fringe signal (or any other source of correlated background noise) a particular instrument would observe.}

We validated using the UBI as a measurement of correlated background noise through a controlled experiment.
We generated images containing only Gaussian noise that were the same size as ZTF images.
The error on each pixel value was set as the root-mean-square of the median-subtracted pixel values in a 10 pixel by 10 pixel box around that pixel.
{\bf The size of this square had to be smaller than the spatial scale of the correlated background noise so that it contained minimal variation in the background value.}
We then laid 50,000 circular apertures onto the image and calculated the aperture flux and aperture flux error of each measurement.
\sout{The UBI was then measured as the ratio of the standard deviation of the aperture fluxes to the median of the aperture flux errors.}
{\bf The UBI as outlined in Equation \ref{eq:UBI} was evaluated five times on each image producing an average UBI as well as an error on that measurement.}

\begin{figure}[!htb]
    \centering
    \includegraphics[width=0.48\textwidth, angle=0]{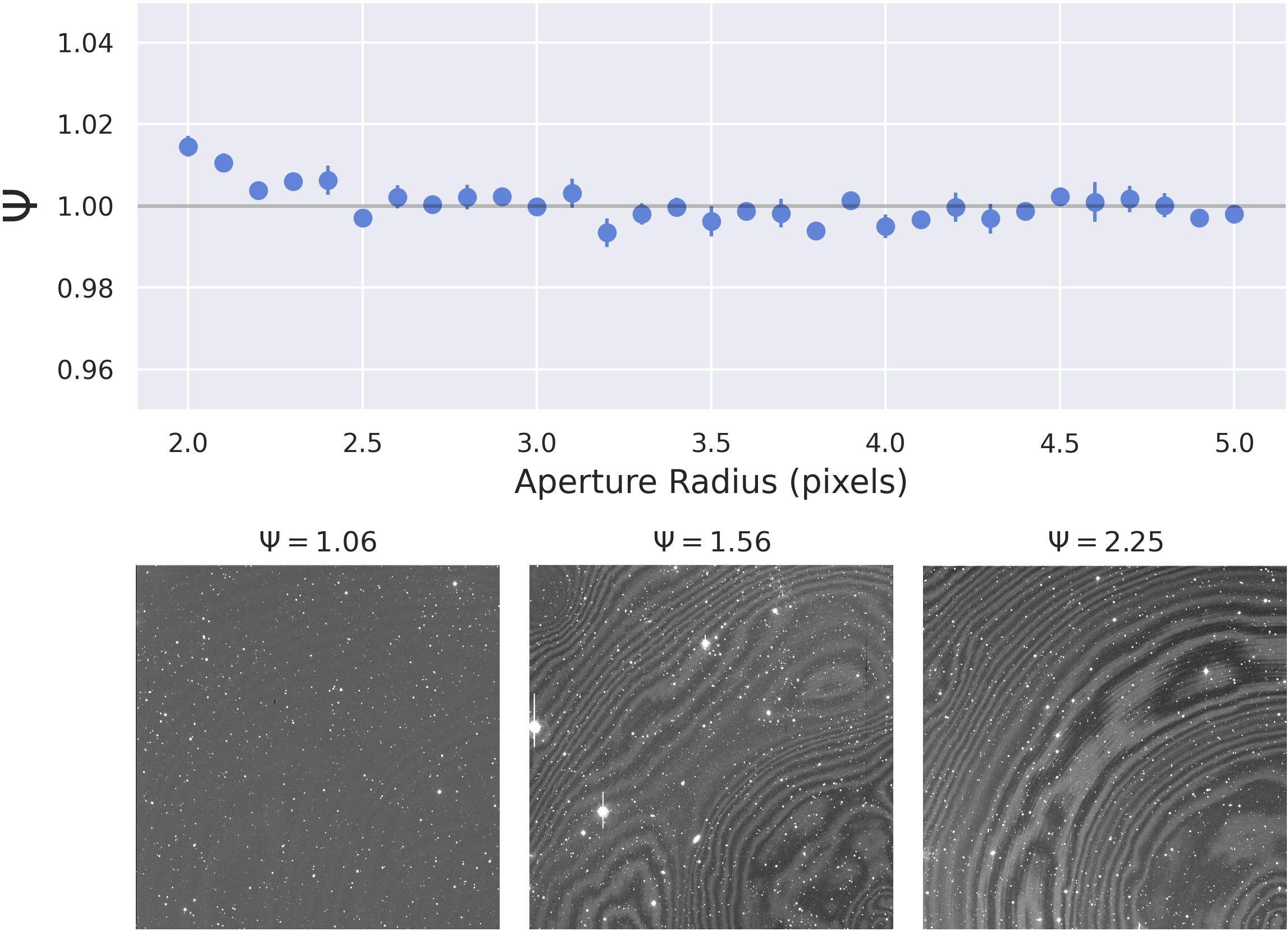}
    \caption{The Uniform Background Indicator is $\Psi\approx1$ for a range of aperture sizes on images of Gaussian noise (top). This validates interpreting $\Psi\approx1$ as measuring no correlated background noise in an image. Three example i-band images and their $\Psi$ values are shown (bottom). More prominent atmospheric fringes correlates with an increase in the value of $\Psi$. \label{fig:UBI_apertures}}
\end{figure}

In Figure \ref{fig:UBI_apertures}, we show the results of this experiment for varying aperture sizes.
For all Gaussian images {\bf and aperture sizes we found} $\Psi\approx1$, confirming our interpretation that this value indicates no correlated background noise on an image.
\sout{After correctly removing the effects of the area weight for the aperture flux errors, this value is consistent for any size of aperture.}
The location and scale of the Gaussian noise was also found to have no effect on this measurement.
Figure \ref{fig:UBI_apertures} shows three example images with correlated background noise and the UBI values that they produce, covering the range of $\Psi$ that we detected throughout this analysis.
These example images show the correlation between a quantitatively larger $\Psi$ and a qualitative increase in the appearance of atmospheric fringes.

Having verified that the UBI is a valid indicator of correlated background noise, we next measured the effect of removing atmospheric fringes on the UBI.
A sample of g-band, r-band and i-band images was created by downloading one random image for each filter, readout-channel and field observed in the ZUDS survey from the week of 2020-02-01 to 2020-02-08 with a limiting magnitude greater than 19.
This sampling method ensured a representative sample of airmasses and limiting magnitude for ZTF observations, while removing images of extremely low quality from the sample.
Our final sample contained 1408 g-band, 1407 r-band, and 1472 i-band images across a range of stellar densities.
Each of the i-band images was cleaned using our method described in Section \ref{sec:method_clean}.
The UBI was then calculated for each of the fringed images, as well as the cleaned images.

\begin{figure}[!htb]
    \centering
    \includegraphics[width=0.48\textwidth, angle=0]{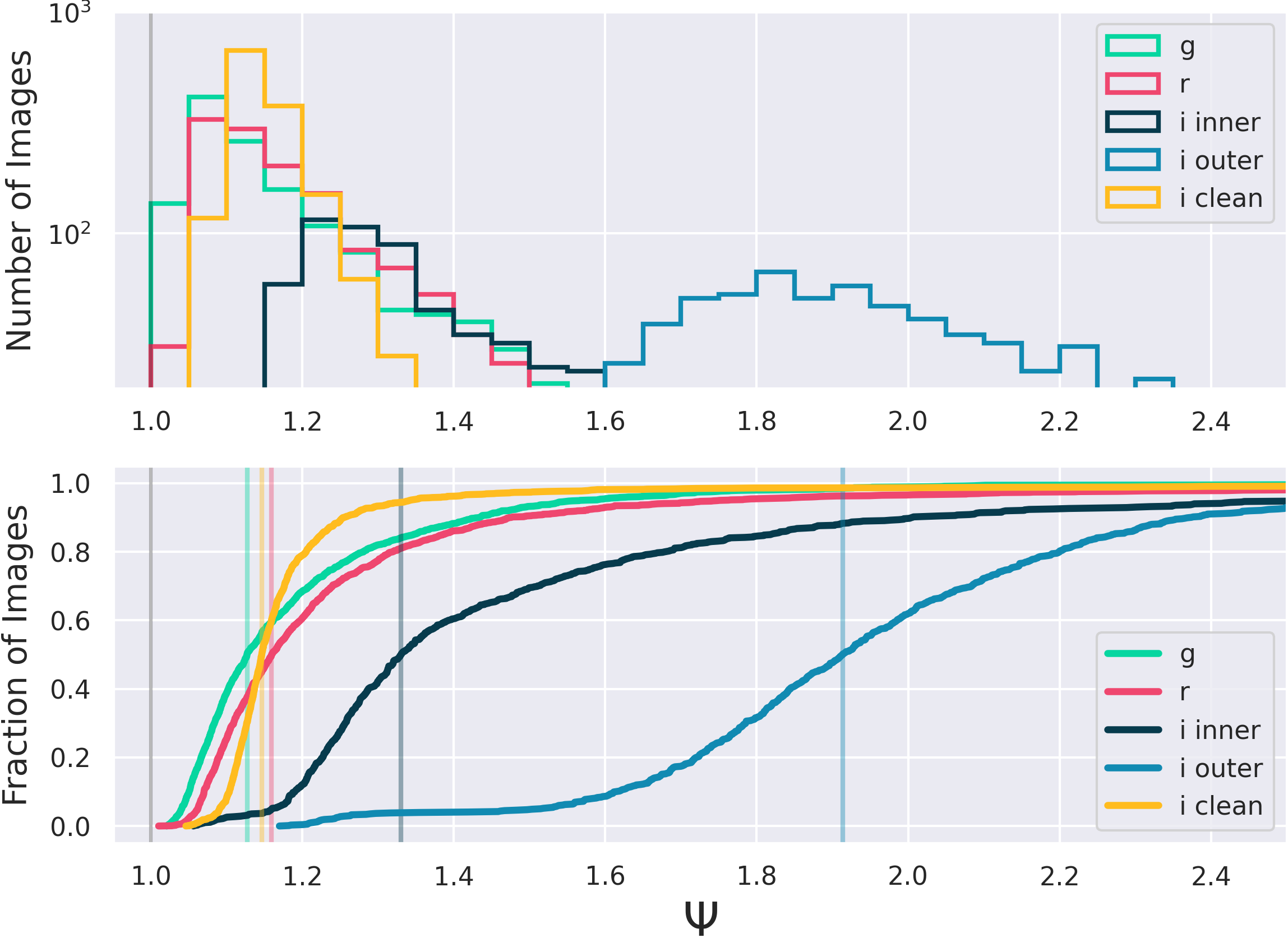}
    \caption{Our sample of g-band, r-band and i-band images show distinctively different distributions in their Uniform Background Indicator ($\Psi$) as shown in both a histogram (top) and cumulative distribution function (bottom). The g-band (green) and r-band (red) average $\Psi\approx1.15$ indicating a small but measurable amount of correlated background noise after flat fielding. The i-band is split between two populations. The images taken on the inner 32 readout channels (light blue) are only moderately affected by atmospheric fringing, averaging $\Psi=1.33$. However the outer 32 readout channels (dark blue) are significantly affected by these fringes, with an average of $\Psi=1.91$ and less than 20\% of the images with $\Psi<1.72$. The population of cleaned i-band images (yellow) is monomodal and has a median value similar to the g-band and r-band of $\Psi=1.15$. ZTF i-band images processed with our method show similar amounts of correlated background noise as is present in g-band and r-band images. \label{fig:UBI_distribution}}
\end{figure}

The distribution of the UBI for all of the images in our sample is shown in Figure \ref{fig:UBI_distribution}, with $\Psi \geq 1$ as expected.
The g-band and r-band images each have a nearly normal UBI distribution with medians of $\Psi=1.13$ and $\Psi=1.16$, indicating a small but measurable amount of correlated background noise remaining after flat-fielding.
The i-band images contain far more correlated background noise and are notably bimodal, with one population averaging $\Psi=1.33$ and a second wider distribution averaging a larger $\Psi=1.91$.
This split is caused by the location of the readout channel on the image plane.
The inner 32 readout channels ($16 \leq \text{rcid} < 48$) with their additional layer of AR coating are less susceptible to atmospheric fringing.
The outer 32 readout channels ($0 \leq \text{rcid} < 16$, $48 \leq \text{rcid} < 64$) experience significantly more fringing due to a lack of an additional later of AR coating.
Less than 20\% of the images have a $\Psi<1.72$ and the 20\% most fringed images have a $\Psi\geq2.20$.
The cleaned i-band images show indistinguishable behavior between the inner and outer readout channels forming a single distribution averaging $\Psi=1.15$.
This population appears similar to the g-band and r-band populations, indicating successful removal of atmospheric fringes.
However the g-band and r-band images have significantly longer tails with an 80th percentile of $\Psi=1.28$ and $\Psi=1.32$ respectively, compared to an 80th percentile for cleaned i-band images of $\Psi=1.20$.
This indicates that there is correlated background noise occurring in g-band and r-band images that PCA analysis could potentially model and remove.
It is clear that the process of removing atmospheric fringes significantly reduces correlated background noise from i-band images.

\begin{figure}[!htb]
    \centering
    \includegraphics[width=0.48\textwidth, angle=0]{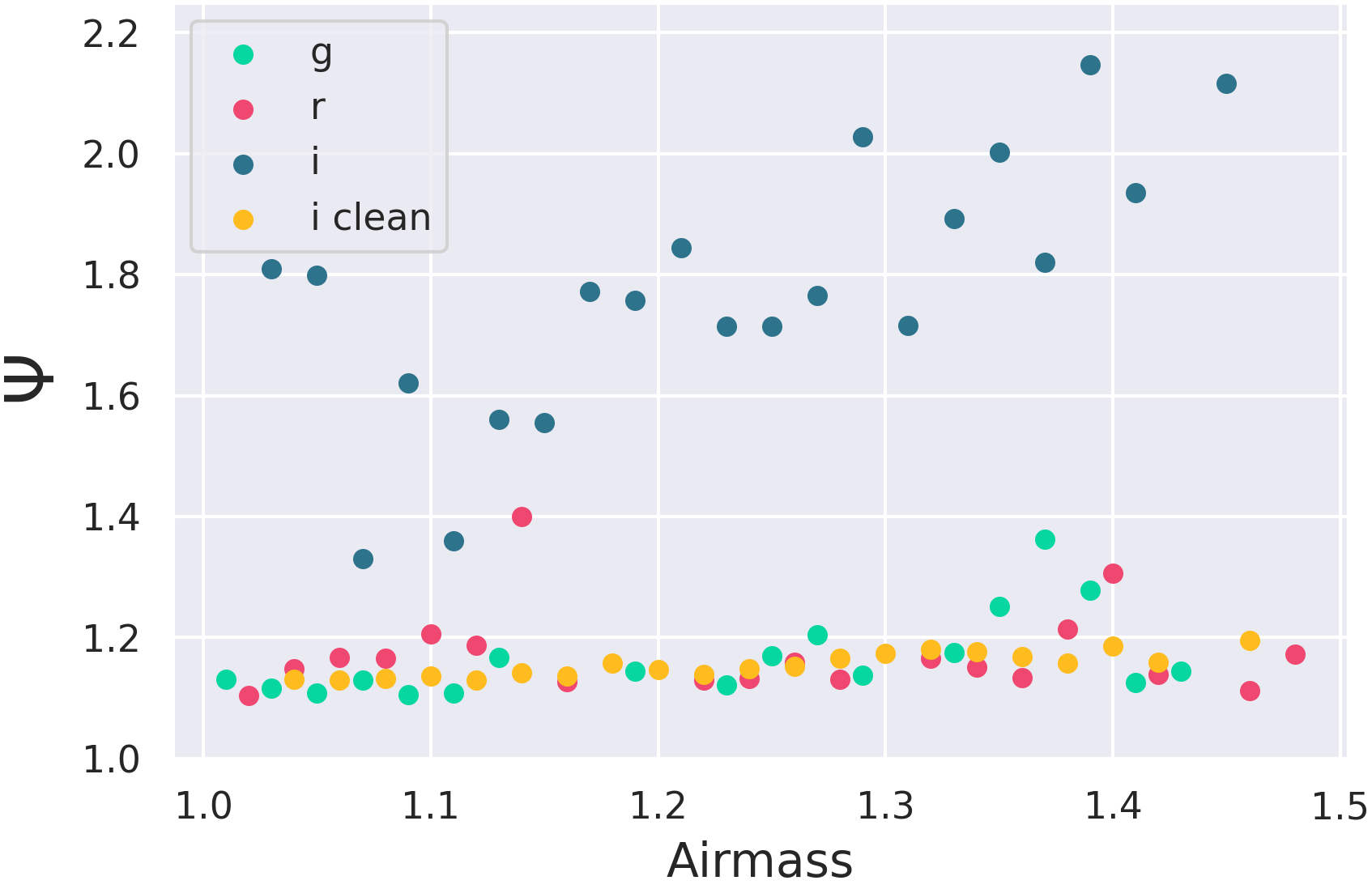}
    \caption{The Uniform Background Indicator ($\Psi$) correlates with airmass for i-band images with significant amounts of atmospheric fringes. Observing sources through the additional column of atmosphere increases the exposure to stimulated emission of atmospheric lines that causes the emergence of fringes. Cleaning the i-band images removes this correlation and produces a relationship with airmass indistinguishable from the g-band and r-band populations. \label{fig:UBI_airmass}}
\end{figure}

It is reasonable to predict that the UBI will increase with airmass, as the presence of atmospheric fringes is caused from the column of atmosphere through which images are taken.
This is found to be the case in Figure \ref{fig:UBI_airmass}, where collecting the images into airmass bins and calculating the median UBI shows a trend toward larger UBI for larger airmass in i-band images.
There is a slight increase in the g-band and r-band images as well, although the effect is relatively weak.
The cleaned i-band images have comparable UBI values to the g-band and r-band images, again confirming the effectiveness of our method. 

\subsection{Photometric Error Due to Fringes}

While the UBI is a useful measurement of the presence of correlated background noise, we sought to quantitatively measure the improvement in photometric precision that resulted from removing atmospheric fringes.
\sout{We will first determine the photometric error caused by atmospheric fringes relative to a reference catalog.
Next we will measure the additional photometric scatter when measuring 5-sigma fake sources injected into a single image.
Last we will measure the recoverability of 0.5-sigma fake sources injected into 100 images before a median co-addition is applied.}
{
\bf
First we measured the photometric error caused by atmospheric fringes relative to a reference catalog.
Next we measured how de-fringing removes this error by injecting faint artificial sources into single images.
Last we measured how de-fringing recovers lost signals by injecting extremely faint sources into images before combining them through co-addition.
}
For each of these experiments we will demonstrate how our method significantly improves the photometric precision of faint source measurements affected by atmospheric fringes.

First we measured the photometric error caused by atmospheric fringes.
For each of the i-band images in our sample, we used SExtractor \citep{sextractor_Bertin1996} to generate an instrumental photometric aperture catalog of astrophysical sources.
We then cross-matched the original and cleaned images with Pan-STARRS1 (PS1) \citep{PanSTARRS} i-band catalogs downloaded from the Vizier database using \texttt{astroquery} \citep{Ginsburg2019} to create cross-matched catalogs.
A zeropoint was calculated for each image using matching sources with PS1 i-band magnitudes less than 17.
This zeropoint was used to transform the instrumental photometric catalogs into ZTF i-band magnitudes comparable with PS1 i-band magnitudes, as well as to calculate a 5-sigma limiting magnitude for each image.
After removing images with a limiting magnitude less than 21 to ensure reasonable measurement of faint sources, our final sample size was 738 images.
For each pair of fringed and cleaned images, the cross-matched catalogs were used to find the variance in the difference between ZTF and PS1 magnitudes for stars within a magnitude bin.
The photometric error caused by fringes was then calculated as:
\begin{align}
    \sigma_{\rm mag} &= \text{std}(m_{\rm PS1} - m_{\rm ZTF}) \\
	\sigma_{\rm fringe} &= \sqrt{\sigma^2_{\rm mag, fringed} - \sigma^2_{\rm mag, cleaned}}
\end{align}

\begin{figure}[!htb]
    \centering
    \includegraphics[width=0.48\textwidth, angle=0]{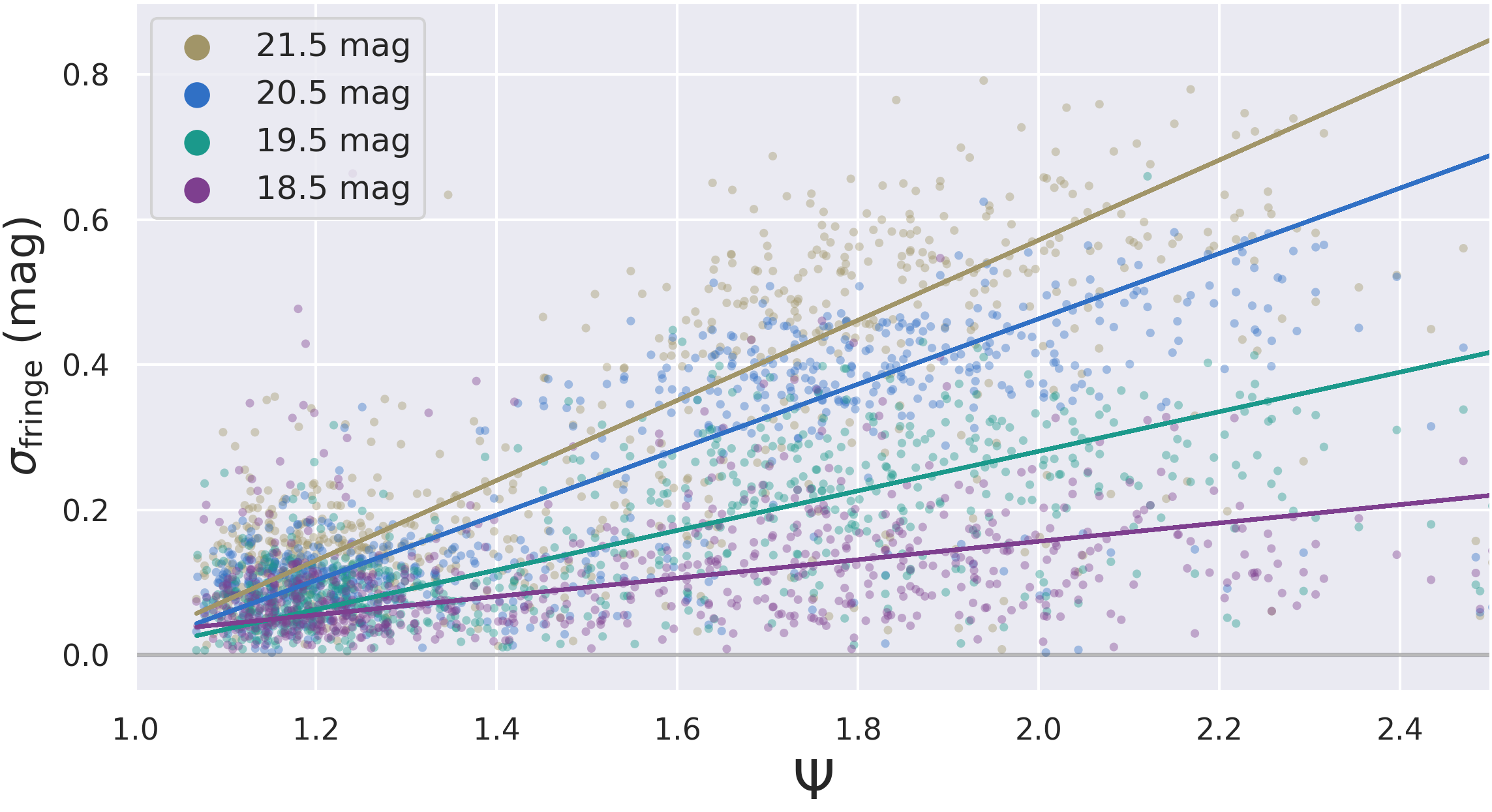}
    \caption{Measurement of the photometric error on faint sources due to fringing on 738 high quality i-band images as shown against the Uniform Background Indicator ($\Psi$) of the fringed image before cleaning. This error is calculated by comparing the variance in the difference between ZTF and PS1 magnitudes before and after removing fringes. Larger UBI values correlate with larger amounts of photometric error, getting as large as 0.46 magnitudes for 20.5 magnitude sources and 0.57 magnitudes for 21.5 magnitude sources at $\Psi=2.0$. Atmospheric fringes add a significant systematic error to the photometry of faint sources that our method is able to remove. \label{fig:UBI_magerr}}
\end{figure}

The photometric error we measured due to fringing is shown in Figure \ref{fig:UBI_magerr}, plotted separately for different faint magnitude bins against the UBI of the fringed images before cleaning.
Images with larger UBI values have larger amounts of additional magnitude scatter relative to the PanSTARRS catalog than those images with smaller UBI values.
Those images with $\Psi \leq 1.3$ appear to have marginal improvement to their photometric precision due to removing fringes.
However, images with $\Psi > 1.3$ show significant photometric errors due to fringing that our method removes.
Images with $\Psi=1.75$ have a 0.21 magnitude error on 19.5 magnitude stars, 0.35 magnitude error on 20.5 magnitude stars and a 0.43 magnitude error on 21.5 magnitude stars.
Images with $\Psi=2.0$ have a 0.28 magnitude error on 19.5 magnitude stars, 0.46 magnitude error on 20.5 magnitude stars and a 0.57 magnitude error on 21.5 magnitude stars.
The worst 10\% of i-band images in the outer readout channels have $\Psi\geq2.4$, resulting in errors as large as 0.21 magnitudes for 18.5 magnitude stars and up to 0.39 magnitudes for 19.5 magnitude stars.
Images with $\Psi\geq2.3$ have very few sources fainter than 19.5 magnitude in the figure, indicating that these images fail to pass the limiting magnitude cut.
\sout{They are a systemic error term in addition to Gaussian noise.}
Our method enables the recovery of the faint end of the luminosity function that would otherwise be unobservable in ZTF i-band images.

\subsection{Effects on Fake Sources: Single Epoch}
\label{sec:fakes_single}

\begin{figure*}[!htb]
    \centering
    \includegraphics[width=0.95\textwidth, angle=0]{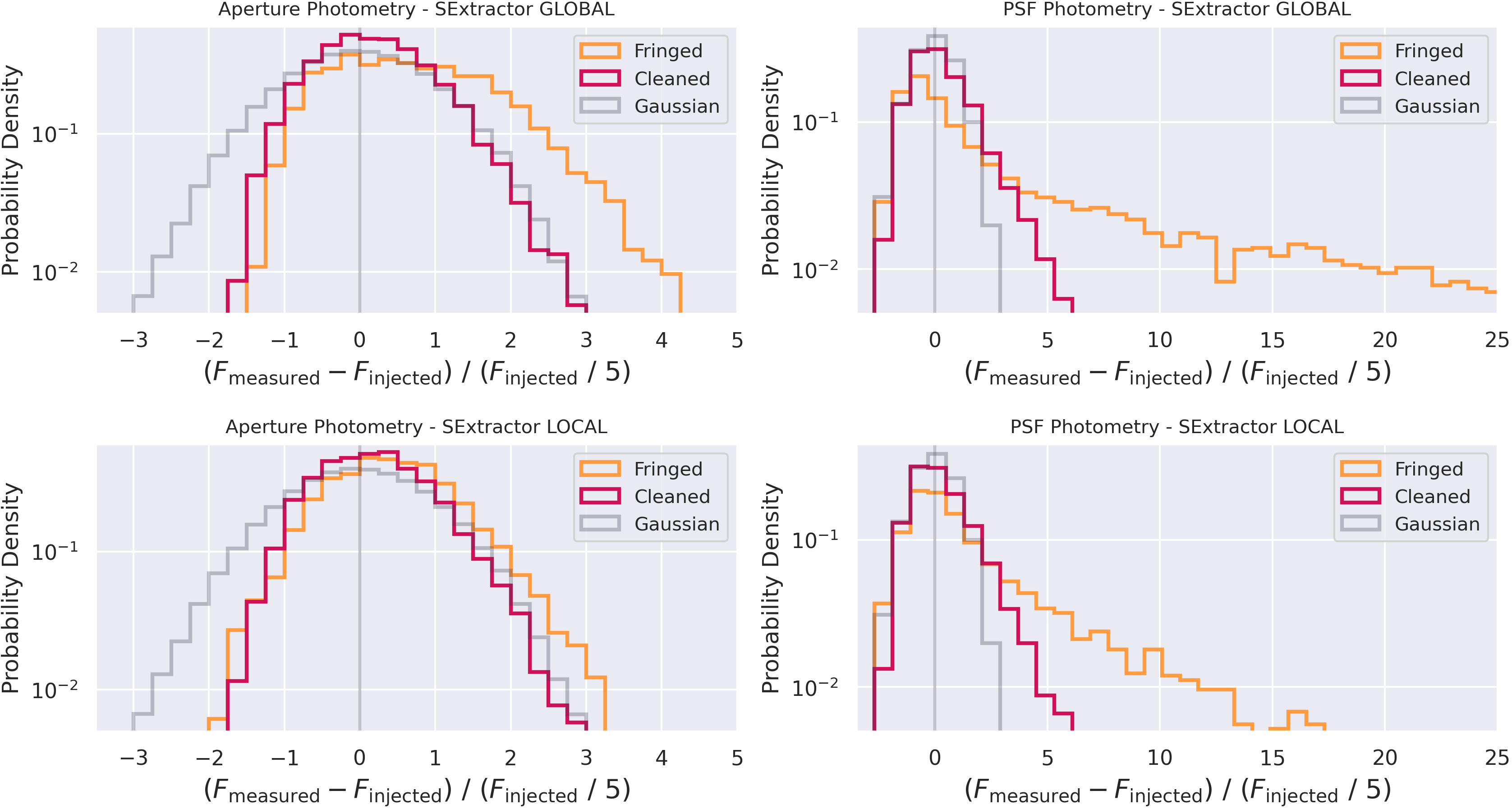}
    \caption{The measured flux of 5000 fake 5-sigma sources injected into an i-band image on the fringed images (orange) and after cleaning (red) using the \texttt{SExtractor GLOBAL} background setting (top), \texttt{LOCAL} background setting (bottom), aperture photometry (left) and PSF photometry (right). On each sub-figure a Gaussian of 5000 sources is drawn (gray) as a visual guide. In all cases the cleaning method significantly increases the accuracy and precision of the recovered flux, particularly for PSF photometry where attempting photometry on images with fringes can often result in overestimating the brightness of the source. Aperture photometry also overestimates the flux resulting in a deficit of lower flux detections than would be statistically expected, although to a lesser degree. If a method for removing fringes cannot be applied, it is best to use the \texttt{SExtractor LOCAL} background setting and a catalog of aperture photometry to most accurately measure the true magnitude of faint sources. \label{fig:fake_phot}}
\end{figure*}

Another way to measure the effect of atmospheric fringes on photometric precision of faint sources is through the injection and recovery of fake sources.
We selected an i-band image with $\Psi=1.78$ as a representative image.
The image's zeropoint and 5-sigma limiting magnitude were calculated using a cross-match to PanSTARRS1 sources as described above.
{\bf A PSF for the image was derived using \texttt{psfex} \citep{psfex} and executed using wrappers written in the \texttt{galsim} \citep{galsim} python package.}
100 sources with magnitudes equal to the 5-sigma limiting magnitude were injected into the original image using this PSF model at random locations, including Poisson noise.
The image was then cleaned using the \texttt{fringez} package.
Photometric catalogs using both aperture and PSF photometry were calculated on the original and cleaned images using \texttt{SExtractor} {\bf with a 3-pixel 1-sigma detection threshold.}
{\bf
Measurements in the catalogs at the location of the injected fake sources were then recovered for comparison to the true modeled flux.
}
We also calculated the ideal aperture corrected flux of the injected fake sources by calculating the median ratio of aperture to PSF fluxes of high signal-to-noise astrophysical sources from the catalogs, for both fringed and cleaned images.
We then repeated this process 50 times for a total of 5000 fake sources injected to form our sample.
We also duplicated this experiment with the \texttt{SExtractor BACKGROUND} parameter set to \texttt{LOCAL} and \texttt{GLOBAL} to test the affect of forcing the source identification algorithm to attempt to characterize local variations in the background noise.
Figure \ref{fig:fake_phot} shows the results of this experiment.
The distributions show the fractional offset of the measured flux to the injected flux scaled by the theoretical error on a 5-sigma source.
Plots are drawn in log-scale to highlight the long tail of overestimated measurements for images that have not been cleaned, with a Gaussian distribution drawn in gray as a reference.

\sout{
We first note some observations about the quality of recovered artificial stars on the images which have not been cleaned.
Setting the \texttt{SExtractor BACKGROUND} parameter to \texttt{LOCAL} has an immediate improvement on making correct measurements for both aperture and PSF photometry, reducing the long tail of overestimating the source's brightness.
This long tail is due to sources that fall onto the additive atmospheric fringes.
\texttt{LOCAL} computes the background flux with a rectangular annulus around the source that prevents attributing the additional brightness in the aperture or PSF model to the source but instead to the background.
PSF photometry is exceptionally poor at correctly identifying this background as a PSF model derived from all sources across the image plane will be artificially broadened on an image with fringes.
Aperture photometry also shows a long tail toward overestimating the brightness due to excessively bright local backgrounds for sources on fringes.
Switching to a \texttt{LOCAL} background also reduces the long tail in this case, showing results that are quite close to those in the cleaned catalog.
It should be noted that for all aperture photometry results there are missing sources at the underestimated flux end of the distribution, as those sources are shifted to the right due to the mis-estimation of the background.
In the absence of a method to remove fringes, performing aperture photometry with a \texttt{LOCAL} background is the best method for minimizing the effects of atmospheric fringes.
}

{
\bf
We first note some observations about the quality of recovered fake sources on the images which have not been cleaned.
Setting the \texttt{SExtractor BACKGROUND} parameter to \texttt{LOCAL} improves photometric measurements on these fringed images.
\texttt{GLOBAL} computes the background flux across the entire image and underestimates the amount of background underneath a source sitting on a bright fringe.
These sources will have overestimated brightness and form the long tail of measurements with larger than expected fluxes.
\texttt{LOCAL} computes the background flux with a rectangular annulus around the source that prevents attributing the additional brightness to the source but instead to the background.
Switching to a \texttt{LOCAL} background therefore reduces this tail.
Aperture photometry performs better than the PSF photometry on these fringed images with a reduced tail for both \texttt{SExtractor} settings.
This prominent tail may be due to a bias in the PSF model caused by including stars that fall on fringes, inflating the wings of the PSF model.
When that model is applied to stars that fall on particularly bright fringes the increased background may be included in the measured flux, resulting in an overestimate of the flux.
Aperture photometry does not suffer from this potential model bias and therefore has a less prominent tail of overestimated fluxes.
}

The sources from the cleaned catalogs are more accurately and precisely measured under all conditions.
{\bf There is a slight tail of overestimated fluxes when performing PSF photometry but it is far closer to a Gaussian distribution than without cleaning.}
For both \texttt{LOCAL} and \texttt{GLOBAL} backgrounds, the distribution of recovered sources in the cleaned catalogs very nearly resembles a Gaussian distribution.
There exists only a few sources with measured fluxes exceeding the injected flux \textbf{causing a deficit of fainter sources.
There still remains a noticeable deficit of sources at the faint end of the brightness distribution, indicating that our method is not removing all additional flux in the image background.}
Aperture photometry remains the best way to evaluate sources even where atmospheric fringes have been removed.
\sout{However the catalogs are invariant to the selection of the \texttt{SExtractor BACKGROUND} parameter.}
Failure to clean images containing atmospheric fringes results in a systematic overestimation of the flux of faint sources.
Applying our method enables near-ideal recovery of 5-sigma sources for a variety of measurement methods.

\subsection{Effects on Fake Sources: Multi Epoch}
\label{sec:fakes_multi}

It is particularly difficult to overcome the effects of fringes when combining multiple images of the same field to recover sources fainter than a single image's limiting magnitude.
Changing atmospheric conditions and various observational airmasses will alter the strength of the fringes, while dithering and sky motion will place astrophysical sources onto slightly different pixels for each  exposures.
Failure to remove fringes will contribute significant excess flux to a co-addition of multiple images.
We demonstrate here this effect quantitatively by injecting extremely faint sources into individual images and attempting to measure them in a co-added image.

100 i-band images of the same field were zero-pointed using the previously outlined method.
{\bf Sources in each image were measured and the signal-to-noise versus magnitude was fit to extrapolate the 0.5 sigma magnitude of all images.}
Each image was injected with {\bf 100 sources at this} 0.5 sigma magnitude \sout{sources} at a common list of coordinates in right ascension and declination.
These sources would be expected to appear as 5-sigma sources after combining the 100 i-band images because signal-to-noise increases as the square root of the total exposure time.
All images were then cleaned with the \texttt{fringez} package.
\texttt{SCAMP} \citep{scamp} was run to find astrometric projection parameters for each of the images such that they could be transformed onto a common reference frame.
All fringed and cleaned images were then combined into two separate co-additions using \texttt{SWarp} \citep{swarp}, produced with a median combination filter and background subtraction.
Aperture photometry catalogs were then generated on the final co-additions using \texttt{SExtractor} with the \texttt{LOCAL} background setting.
Sources at the locations of the injected signals were recovered and their signal-to-noise measured as the ratio of their aperture flux to their locally determined aperture flux error.
We repeated this process 50 times for a total of 5000 fake sources distributed over 50 fringed and 50 cleaned co-additions.

\begin{figure}[!htb]
    \centering
    \includegraphics[width=0.48\textwidth, angle=0]{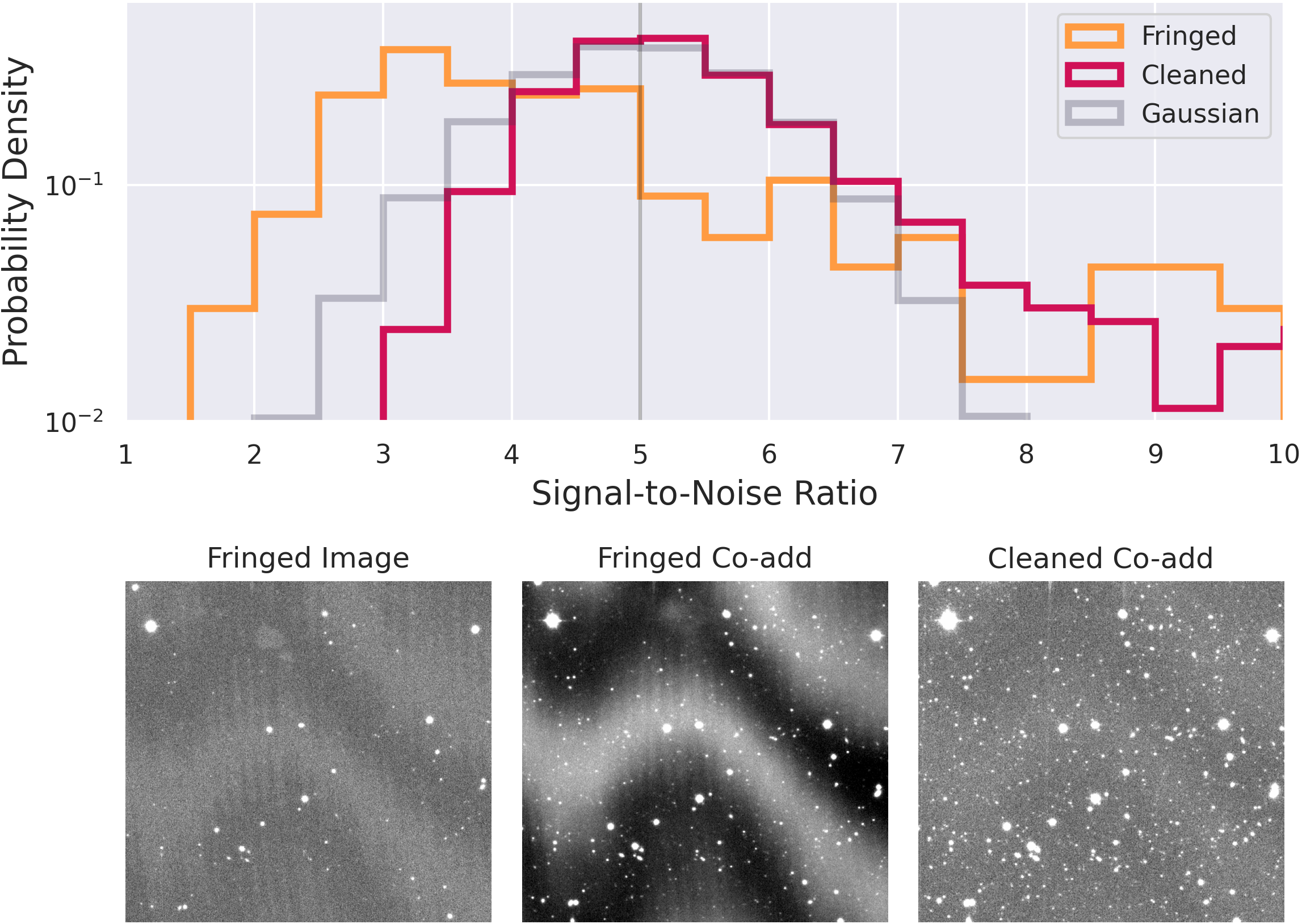}
    \caption{The signal-to-noise distribution of 5000 0.5 sigma sources injected into 100 images after a median co-addition (top) for images with atmospheric fringes (orange) and after cleaning (red). The photometric catalogs were generated with a \texttt{SExtractor LOCAL} background and aperture photometry. 1161 sources were recovered in the cleaned images and only 156 sources were recovered in the fringed images, demonstrating the necessity to clean i-band images in any attempt to find faint sources after multi-epoch co-addition. Those sources that were recovered in the fringed images are most likely to be detected at an artificially low signal-to-noise, with a long tail of higher signal-to-noise due to falling on positive fringes. The cleaned images have a distribution much closer to Gaussian, with a small deficit at the low signal-to-noise end that also appears as a tail at the higher end. The photometric improvement in multi-epoch co-additions can be clearly seen in as a smoothly varying background and the presence of faint sources after cleaning has been applied (bottom). \label{fig:fake_phot_coadd}}
\end{figure}

Figure \ref{fig:fake_phot_coadd} shows the quantitative and qualitative photometric improvements to these recovered sources.
Sources observed on fringed co-additions peak at a signal to noise of 3, pushing them below the typical 5-sigma observable threshold.
These sources also have a long tail of excessive flux stretching as high as 10 sigma due to the additional flux caused by atmospheric fringes.
Sources observed on cleaned co-additions are much closer to their theoretical distribution.
These sources peak at exactly 5-sigma and the large majority of sources are within the normal signal-to-noise range of 4 to 6.
There still exists a tail of excessive fluxes, demonstrating that our method is failing to remove all excessive correlated background from the individual exposures.
However our method clearly produces an observed flux distribution that is much closer to a Gaussian distribution.

The largest difference between the two populations appears in their yields.
Of the 5000 sources injected into each of the fringed and cleaned images, 1161 sources were recovered from cleaned co-additions and only 156 sources were recovered from fringed co-additions.
Over 95\% of the 0.5 sources injected into the fringed images failed to be recovered after co-addition.
The order of magnitude increase in the number of sources recovered by our method enables ZTF i-band surveys to recover faint sources that would otherwise have been extremely unlikely to observe.
We note that forced photometry at the known locations of the fake sources may have increased the recovered yields for both populations but would not be an accurate representation of the observation process undertaken for unknown sources.

Our analysis demonstrates the power of our method to remove photometric error due to atmospheric fringes and enable the recovery of faint sources in both single images and co-additions that would otherwise have been undetectable.
Failure to implement a method for removing atmospheric fringes greatly reduces the effectiveness of the i-band filter for observing any sources fainter than 18th magnitude.
Our method increases the photometric precision of the i-band to that of g-band and r-band images that do not suffer from atmospheric fringes.

\section{Discussion} \label{sec:conclusion}

Our method has several benefits beyond improving the photometric quality of individual i-band images.
Image references are constructed through the combination of as many as 40 epochs taken at a single field.
As shown in Section \ref{sec:fakes_multi}, removing fringes significantly improves the photometric precision on a multi-epoch co-addition.
Application of our method also greatly improves the quality of the ZTF alert stream.
The alert stream packets are generated on difference images created from subtracting individual epochs from a multi-epoch co-addition reference (after appropriate scaling).
Correctly removing fringes makes visible the faint end of the luminosity function that would otherwise not be observable in the i-band alert stream.

Future work on removing atmospheric fringes using PCA could improve upon our method in several ways.
Training on CCDs instead of readout channels could produce better eigen-images by including more correlated pixels in the PCA feature identification.
Also, different methods of calculating the local flux error such as calculating the root-mean-square error on a smaller (or larger) square around each pixel could produce a less noisy UBI measurement.
Lastly, investigating the eigenvalues for fringe bias images generated on different readout channels from the same exposure could reveal correlations that could be used to perturb the fringe bias into a better fit for each readout channel.
Treating each readout channel as entirely independent, while convenient and a natural fit for the IPAC processing pipeline, may leave out valuable information that could improve our method.

The eigen-images shown in Figure \ref{fig:eigenimages}, as well as the eigen-images of many of the other fringe models, show significant smooth variations that are not being removed by the current flat fielding pipeline.
This indicates that the fringe bias images include not only atmospheric fringes, but residual global gradients.
Future work could be done on exploring the application of this PCA method to supplement or even replace the current flat fielding pipeline on not only the i-band images, but g-band and r-band images as well.
The Uniform Background Indicator can be used as a quantitative measurement to compare how well a PCA method, as compared to more classical flat fielding methods, generates astronomical images with normal backgrounds.

\begin{acknowledgments}
This work is based on observations obtained with the Samuel Oschin Telescope 48-inch and the 60-inch Telescope at the Palomar Observatory as part of the Zwicky Transient Facility project.
ZTF is supported by the National Science Foundation under Grant No. AST-1440341 and a collaboration including Caltech, IPAC, the Weizmann Institute for Science, the Oskar Klein Center at Stockholm University, the University of Maryland, the University of Washington, Deutsches Elektronen-Synchrotron and Humboldt University,  Los Alamos National Laboratories, the TANGO Consortium of Taiwan, the University of Wisconsin at Milwaukee, and Lawrence Berkeley National Laboratories.
Operations are conducted by COO, IPAC, and UW.

We acknowledge support from the University of California Office of the President for the UC Laboratory Fees Research Program In-Residence Graduate Fellowship (Grant ID: LGF-19-600357).
This research used resources of the National Energy Research Scientific Computing Center, a DOE Office of Science User Facility supported by the Office of Science of the U.S. Department of Energy under Contract No. DE-AC02-05CH11231.
We acknowledge support from the DOE under grant DE-AC02-05CH11231, Analytical Modeling for Extreme-Scale Computing Environments.
M.~W.~Coughlin acknowledges support from the National Science Foundation with grant number PHY-2010970.

\end{acknowledgments}

\bibliographystyle{aasjournal}
\bibliography{references.bib}

\end{document}